\documentstyle[emulateapj,psfig]{article}

\begin{document}

\title{Identification of Two categories of optically bright
$\gamma$-ray bursts}

\author{Enwei Liang$^{1,2}$ and Bing Zhang$^{1}$}
\affil{$^1$Department of Physics, University of Nevada, Las Vegas, NV
89154, USA\\
Email:lew@physics.unlv.edu; bzhang@physics.unlv.edu\\ $^2$Physics
Department, Guangxi University, Nanning 530004, P. R. China}

\begin{abstract}
We present the results of a systematical analysis of the intrinsic
optical afterglow light curves for a complete sample of gamma-ray
bursts (GRBs) observed in the period from Feb. 1997 to Aug. 2005.
These light curves are generally well-sampled, with at least four
detections in the $R$ band. The redshifts of all the bursts in the
sample are available.  We derive the intrinsic $R$ band afterglow
lightcurves (luminosity versus time within the cosmic proper rest
frame) for these GRBs, and discover a fact that they essentially
follow two universal tracks after 2 hours since the GRB triggers. The
optical luminosities at 1 day show a clear bimodal distribution,
peaking at $1.4\times 10^{46} ~{\rm ergs~s^{-1}}$ for the luminous
group and $5.3\times 10^{44} ~{\rm ergs~s^{-1}}$ for the dim
group. About $75\%$ of the GRBs are in the luminous group, and the
other $25\%$ belong to the dim group. While the luminous group has a
wide range of redshift distribution, the bursts in the dim group all
appear at a redshift lower than 1.1.

\end{abstract}

\section{Introduction} Gamma-ray bursts (GRBs) are believed to be the
brightest electromagnetic explosions in the universe after the
identification of their cosmic origin (Metzger et al. 1997). Two
categories of these erratic, transient events have been identified,
i.e. long-soft and short-hard (Kouveliotou et al. 1993). The
association of long GRBs with very energetic core-collapse supernovae
has now been well established (Galama et al. 1998; MacFadyen et
al. 1999; Bloom et al. 1999; Stanek et al.  2003; Hjorth et al. 2003;
Thomsen et al. 2004; Malesani et al. 2004). Several short GRBs have
been localized and observed by Swift and HETE-2 recently, which are
found to reside in nearby galaxies, some of which are of early-type
with little star formation (Gehrels et al. 2005; Fox et al. 2005;
Villasenor et al. 2005; Hjorth et al. 2005a; Barthelmy et al. 2005;
Berger et al. 2005). This indicates that they have a distinct origin
from the long species. Most of the well localized GRBs, both long and
short, are followed by long-lived, decaying afterglows in longer
wavelengths (Costa et al. 97; van Paradijs et al. 1997; Frail et
al. 1997; Gehrels et al. 2005; Fox et al. 2005). Long GRBs have been
themselves classified into two groups, optically bright and optically
dark, based on whether or not an optical transient is detected to a
given brightness limit at a given time delay (e.g. Groot et al. 1998;
Fynbo et al. 2001; Berger et al. 2002; Jacobsson et al. 2004; Rol et
al. 2005). The origin of optically dark GRBs is still unclear. Very
early, tight upper limits made by the Swift UV-Optical Telescope
indicate that the darkness is not caused by observational biases
(Roming et al. 2005). Based on X-ray afterglow data, a tentative
bimodal distribution of X-ray luminosities has been also noticed
(B\"oer \& Gendre 2000; Gendre \& B\"oer 2005).

Over more than 8 years of optical afterglow hunting, more than 70
optically-bright GRBs have been detected, among which 44 bursts have
well-sampled light curves and redshift measurements (\S 2). In this
Letter we present a systematical analysis to these 44 optical
afterglow light curves in the cosmic rest frame. We find a fact that
their late-time lightcurves follow two apparent universal tracks (\S
3). We then conclude that within the optically bright GRBs there exist
two sub-categories, the luminous group and the dim group (\S
4). Cosmological parameters $\Omega_M=0.3$, $\Omega_\Lambda=0.7$, and
$H_0=71$ km Mpc$^{-1}$ s$^{-1}$ have been adopted throughout this
Letter.

\section{Data}
We make a complete search from the literature for the $R$-band
afterglow light curves detected during the time period from Feb. 1997
to Aug. 2005. We obtain a GRB sample with 44 GRBs, which is tabulated
in Table 1\footnote{A full version of the GRB sample with references
to the observational data are available in the electronic
version}. These light curves have at least four detections in the
$R$-band. The redshifts of the bursts are available. We collect the
following data for these bursts from published papers or from GCN
reports if the former are not available\footnote{We collect the
$\beta$ and the extinction $A_{V}$ of each burst from the same
literature to reduce the uncertainties introduced by different
authors.}, i.e. redshift ($z$), $R$-band magnitude, spectral index
($\beta$), and extinction by the host galaxy ($A_V$). For those bursts
whose $\beta$ and $A_V$ are not available, we take $\beta=0.75$, the
mean value of $\beta$ in our sample, and $A_V=0$.  Galactic extinction
correction is made by using a reddening map presented by Schlegel et
al. (1998). The extinction curve of the Milky Way\footnote{We also
tried other types of extinction curves, and found that our results are
insensitive to the extinction model adopted.}(Pei 1992) is adopted to
calculate the extinction in the local frame of the GRB host galaxy.
The $k$-correction in magnitude is calculated by
$k=-2.5(\beta-1)\log(1+z)$. For late time data, possible flux
contribution from the host galaxy is subtracted.

\section{The Bimodal Luminosity Evolutions}
We convert the corrected magnitudes to fluxes ($F^{c}$) by using the
photometric zero points given by Fukugita et al. (1995). The
luminosity at the cosmic proper time $t^{'}$, $L_R(t')$, is calculated
by $L_R(t')=4\pi D_L^2(z) F^{c}$, where $D_L(z)$ is the luminosity
distance at $z$. The luminosity error is calculated by $ \Delta \log
L_R=\{0.16(\Delta R^2 +\Delta A^2_{R^{'}})+[\Delta \beta
\log(1+z)]^2\}^{1/2}$, where $\Delta R$ is the observed uncertainty of
the $R$ band magnitude, $\Delta A_{R^{'}}$ is the uncertainty of the
host galaxy extinction at the cosmic rest frame wavelength
$\lambda_{R^{'}}=\lambda_R/(1+z)$, and $[\Delta \beta \log(1+z)]$ is
the error of the $k$-correction.

The intrinsic $R$-band light curves [$L_R(t')$ vs. $t'$] are displayed in Figure 1 for 42
bursts. The two nearby GRBs, 980425 and 031203 are not included, since their light curves
are significantly contaminated by the underlying supernova component (Galama et al. 1998;
Thomsen et al. 2004). It is found that although the light curves at $t^{'}<0.1$ days vary
significantly, they are clustered and follow two apparent universal tracks at $t' > 0.1$
days, indicating that within the optically bright GRBs there exist two well-separated
sub-categories. The majority of the bursts ($\sim 75\%$) comprises an optically luminous
GRB group, which includes the well-studied GRBs such as 030329, 990123, and 990510. It is
interesting that although the isotropic gamma-ray energy ($E_{\gamma, \rm iso}$) of GRB
990123 and GRB 030329 differ by almost 2 orders of magnitude, their late optical
afterglow luminosities are similar\footnote{We notice that Nardini et al. (2005)
independently obtained the same result during the process when our paper was being
reviewed.}. The other $\sim 25\%$ GRBs in our GRB sample comprises the dim group, with
the representative bursts being GRBs 021211 and 041006. We zoom in these light curves in
the time regime from 0.1 days to 10 days in the inset of Figure 1. The bimodal lightcurve
trajectories during this are more clearly visible. Based on the separation of the two
groups by the luminosity at $1$ day ($\log L_{R,1d}/{\rm erg\ cm^{-2}}=45.15$, see Figure
2) and adopting a typical temporal decay index $\sim -1.2$, we draw a division line for
the two groups as $\log L_R=45.15-1.2\log t^{'}$ (the dashed line in Figure 1). It is
found that 25 (out of 34) and 7 (out of 10) light curves in the luminous and dim groups,
respectively, cover this time regime and do not cross over the division line. They are
the most representative (with the smallest scatter) ones in both groups. The bursts in
the luminous group are typically brighter than those in the dim group by a factor of
$\sim 30$.

We read off or extrapolate/interpolate the luminosity at a given epoch
from the light curves, and perform rigorous statistics to access the
bimodality of our sample. We first select the intrinsic luminosity at
1 day for our purpose. Our consideration is two folds.  First, the
early optical light curves may have contributions from the reverse
shock component or additional energy injection from the central
engine. The optical band may be below the cooling frequency or even
below the typical synchrotron frequency so that the flux sensitively
depends on many unknown shock parameters. On the other hand, the late
emission is fainter and may contain luminosity contamination from the
host galaxy.  Second, most of the observations were made around this
epoch. This makes the luminosity derivations more reliable. Figure 2
shows the 2-dimensional distribution of the intrinsic R-band
luminosity at 1 day\footnote{In view of the difficulty of subtracting
the supernova contribution from GRB 980425 (Galama et al. 1998) and
GRB 031203 (Thomsen et al. 2004), we use the first two data points
(which are around 1 day) in each burst's light curve to derive the
upper limits of their luminosities at $1$ day, both giving $\sim
7\times 10^{43} ~{\rm erg ~s^{-1}}$. The Galactic extinction corrected
luminosities are $8.3\times 10^{43} ~{\rm erg ~s^{-1}}$ for GRB 980425
and $9.2\times 10^{44} ~{\rm erg ~s^{-1}}$ for GRB 031203.},
$L_{R,1d}$, versus $E_{\gamma, \rm iso}$ (panel a), and the
distributions of the two quantities, respectively (panels b and
c). Flux thresholds in both the $\gamma$-ray and the optical bands
introduce selection effects against low-energy, low-luminosity bursts,
and these are indicatively marked as the grey regions in Figure
2. There are three most prominent outliers whose light curves deviate
from the universal light curves, i.e. GRBs 970508, 030226, and
050408. They are excluded in the statistical analyses (see more
detailed discussion in \S 4). While the $E_{\gamma,\rm iso}$
distribution displays a power-law with sharp cutoff around $10^{51.5}$
ergs (due to the selection effect), $\log L_{R, 1d}$ shows a
well-defined bimodal distribution, which is well fitted by a two
Gaussian model centered at $\log L_{c,1}/~{\rm 1 \ erg~s^{-1}}=44.66$
with $\sigma_1=0.41$ and $\log L_{c,2}/~{\rm 1 \ erg ~s^{-1}}=46.15$
with $\sigma_2=0.77$. The bimodality is at a confidence level of
3$\sigma$ tested by a classification algorithm with the minimum
Euclidian distance discriminant and the KMM algorithm (Ashman et
al. 1994). A bootstrap test ($10^5$ bootstrap samples) shows that the
distributions of the means of $\log L_{R, 1d}$ of the two groups and
their covariance ($c$) are normal, which gives $\log L_{c,1}/~{\rm 1 \
erg~s^{-1}}=44.72^{+0.36}_{-0.36}$, $\log L_{c,2}/~{\rm 1 \
erg~s^{-1}}=46.15^{+0.14}_{-0.20}$, and $c=0.11^{+0.16}_{-0.06}$ at
$3\sigma$ significance level. These results indicate that the
bimodality is not due to statistical fluctuations.

In order to further examine the bimodal distribution at different
epoches, we also derive the distributions at $\log t^{'}/ {\rm 1\
day}=-0.5$ and 0.5, respectively. We find that the distribution of the
luminosities at $\log t^{'}/{\rm 1\ day}= 0.5$ is bimodal with a
$3\sigma$ significance level. The bimodality of the luminosity
distribution at $\log t^{'}/{1\ \rm day}=-0.5$ has a lower
(i.e. $2\sigma$) statistical significance. Nonetheless, the
distribution still stands with a gap at $\log L_R/{\rm erg\
s}^{-1}=45.5$. The lower significance is expected, because of the
various factors (e.g. reverse shock, early injection, etc) concerning
the early afterglows.

\section{Conclusions and Discussion}
We have derived the intrinsic $R$ band afterglow lightcurves within
the cosmic proper rest frame with a completed sample observed from
Feb. 1997 to Aug. 2005. These light curves follow two apparent
universal tracks after 2 hours since the GRB triggers. The optical
luminosity at 1 day clearly shows a bimodal distribution, with the
peak luminosities being $1.4\times 10^{46} ~{\rm ergs~s^{-1}}$ for the
luminous group and $5.3\times 10^{44} ~{\rm ergs~s^{-1}}$ for the dim
group.

One interesting feature for the dim group is that these bursts all
appear to have low redshifts. It has been previously speculated that
nearby GRBs might be different from their cosmological brethren
(Norris 2002; Soderberg et al. 2004; Guetta et al. 2004). In our
sample, the two well-known nearby GRBs, 980425 and 031203, both belong
to the dim group. Except GRB 980613 ($z=1.096$) and GRB 021211
($z=1.006$), other bursts in the dim group all have $z<1$. Besides the
low-$z$ property, the bursts in the dim group all have an isotropic
$\gamma$-ray energy much lower than that of the bursts in the luminous
group. They also have simple lightcurves. All the bursts in the dim
group have a single gamma-ray pulse, except for GRB 990712 who has two
well-separated pulses. We notice that the observed $R$-band magnitudes
for the dim GRBs are generally $\sim (21-22.5)$ mag a few days after
the trigger. Although a burst with $\log (L_{R}/{\rm erg \
s^{-1}})=44.72$ (the typical 1-day optical luminosity for the dim
group) should be detected up to $z=2.4$ for an observation threshold
of $R\sim 22.5$ mag, the efficiency to detect optical transients
fainter than $R\sim 21$ is dramatically reduced. The observational
bias for the deficit of high-redshift, optical-dim GRBs thus cannot be
ruled out.

The extinction effects have been carefully taken into account. The
data indicate that the dim GRBs do not exhibit significantly higher
extinction than the luminous ones. It has been suggested that dust in
the host galaxy may be destroyed by early radiation from $\gamma$-ray
bursts and their afterglows (Waxman et al. 2000; Fruchter et
al. 2001). It is found that the optical extinctions are $10-100$ times
smaller than what are expected from the X-ray absorption (Galama et
al. 2001), and that the dimness of GRB 021211, a representative burst
in our dim group, could not be explained by the extinction effect
(Holland et al. 2004). The apparent bimodality therefore could not be
interpreted by the extinction effect. Our results then suggest that
there might be two types of progenitors or two types of explosion
mechanisms in operation.

Some GRBs show an initial shallow decay before landing onto the
luminous branch. GRB 970508 is the most prominent one. The light
curve is initially almost flat before re-brightening at about 0.5
days, peaks at 1 day, and eventually settles
onto the luminous branch, although with significant fluctuations
(Pedersen et al. 1998). These fluctuations are similar to those
observed in GRBs 000301C, 021004, and 030329. The initial shallow
decay and fluctuations are thought to be due to additional energy
injections during the afterglow phase (Dai \& Lu 2001; Bj\"ornsson et
al.  2004; Fox et al. 2003; Zhang et al. 2005). GRBs 050408 and 050319
have the similar behavior. When injection is essentially over, the
total afterglow kinetic energies of these bursts are similar to those
of the bursts in the luminous group. Therefore they should be
classified into the luminous group. Another type of outliers are those
light curves with a sharp rapid decay at early times. GRB 030226 is
the most prominent one in our sample. This may be attributed by an
early jet break, and the rapid decay effect is due to the sideways
expansion of the jet, which significantly reduces the optical
luminosity (Rhoads 1999).

The two apparent universal lightcurve tracks at later times are
intriguing. It is widely believed that afterglows are synchrotron
emission from shocked circumburst medium as the fireball is
decelerated (M\'esz\'aros \& Rees 1997; Sari et al. 1998; see also
reviews by M\'esz\'aros 2002, Zhang \& M\'esz\'aros 2004, Piran
2005). At a late enough epoch, the optical band may be above both the
typical synchrotron frequency and the synchrotron cooling
frequency. In such a spectral regime and at a particular epoch
(e.g. $t' =1$ d), the optical luminosity $L_{R,1d} \propto E_{\rm k,
iso}^{(p+2)/4} \epsilon_e^{p-1}
\epsilon_B^{(p-2)/4}$, where $E_{\rm k,iso}$ is the isotropic kinetic energy of the
fireball, $\epsilon_e$ and $\epsilon_B$ are shock energy equipartition
factors for electrons and magnetic fields, respectively, and $p$ is
the electron spectral index. We can see that $L_{R,1d}$ is
medium-density-independent, and only weakly depends on
$\epsilon_B$. The universal afterglow luminosity therefore suggests
that both $E_{\rm k,iso}$ and $\epsilon_e$ are standard values around
1 day for each subclass. A standard $\epsilon_e$ suggests universal
properties of relativistic shocks. A standard $E_{\rm k,iso}$, on the
other hand, is intriguing, since $E_{\gamma,\rm iso}$ vary for 4
orders of magnitude among long duration GRBs and they generally follow
a power-law distribution with a cutoff at low luminosity end (Schmidt
2001, Norris 2002). They become standard only when jet beaming
correction is taken into account (Frail et al. 2001). Our results are
consistent with the picture that GRBs with a higher $E_{\gamma,\rm
iso}$ tends to have a higher $\gamma$-ray emission efficiency
(Lloyd-Ronning et al. 2004). The $E_{\rm k,iso}$ derived using 10-hour
X-ray data requires a jet beaming correction to achieve a standard
value (Berger et al. 2003). The early X-ray afterglows in the cosmic
proper frame for a group of GRBs observed with the {\em Swift} X-Ray
Telescope indicate a large scatter of $E_{\rm k,iso}$ at early time
(Chincarini et al. 2005). Our results therefore suggest a possible
evolution of $E_{\rm k,iso}$ with time. One scheme might be that GRB
jets are initially structured (Zhang \& M\'esz\'aros 2002; Rossi et
al. 2002), and the early $\gamma$-ray and X-ray properties are
sensitive to the observer's viewing angle.  The jet structure tends to
smear out with time, so that at later times, the outflow is more
isotropic and the viewing angle effect no longer plays an essential
role.

We appreciate constructive comments from the referees during the
reviewing process of this paper both in ApJ Letters and Nature. This
work is supported by NASA under grants NNG05GB67G, NNG05GH92G, and
NNG05GH91G, as well as the National Natural Science Foundation of
China (No. 10463001).

\begin{table}[t]
\small

 \caption[]{The GRB sample with well-sampled optical afterglow light curves and
known redshifts}
\begin{tabular}{llllllllllllllllll}

\hline\hline GRB$^{a}$& $z$ & $\beta(\Delta \beta)$& $A_{\rm V, host}(\Delta A_{\rm V,
host})$&GRB$^{a}$& $z$ & $\beta(\Delta \beta)$& $A_{\rm V, host}(\Delta A_{\rm V,
host})$\\
\hline

970228&0.695&0.780(0.022)  &0.5 &
970508&0.835 &1.11  &0\\

971214&3.42&0.87(0.13) &0.43 (0.08)  &
{\bf 980326}& 1.0 &0.8(0.4) &0\\

{\bf 980425}&0.0085 &-&-&
{\bf 980613}&1.096 &0.60&0.45\\

980703& 0.966&1.013 (0.016)&1.50 (0.11)&
990123& 1.6004 &0.750 (0.068)&0\\

990510&1.6187 &0.55 &0&
{\bf 990712}&0.434&0.99 (0.02) &0 \\

991208& 0.706&0.75 &0 &
991216& 1.02 &0.60 &0\\

000131& 4.5&0.70 &0.18 &
000301C& 2.03 &0.70 &0.09 \\

000418& 1.118&0.75  &0.96 &
000911& 1.058&0.724(0.006) &0.39 \\

000926& 2.066&1.00(0.18)  &0.18(0.06) &
010222& 1.477&1.07 (0.09)  &0 \\

{\bf 011121}& 0.36&0.80(0.15)  &0  &
011211& 2.14&0.56(0.19)&0.08(0.08)  \\

020124& 3.198&0.91 (0.14) &0 &
020405& 0.69&1.43(0.08) &0\\

020813& 1.25&0.85(0.07)  &0.14(0.04)&
{\bf 020903}&0.25&-&-\\

021004& 2.335 &0.39 &0.3&
{\bf 021211}& 1.01 &0.69 &0 \\

030226& 1.98&0.70(0.03) &0  &
030323& 3.372&0.89(0.04) &$<0.5$  \\

030328& 1.52&-&- &
030329& 0.17&0.5  &0.30(0.03)  \\

030429& 2.65&0.75 &0.34 &
030723&2.10&1.0&0.4\\

{\bf 031203}& 0.105&-&- &
{\bf 040924}& 0.859&0.70 (0)  \\

{\bf 041006}& 0.716&0.55 &0&
050315& 1.949&-&-\\

050319& 3.24&-&- &
050401& 2.90 &-&-\\

050408& 1.24&-&-  &
050502& 3.793& -&-  \\

050525& 0.606&0.97(0.10) &0.25(0.16) &

050603&2.821&-&- \\
050730&3.97&-&-&

050820&2.615&-&-&
&&-&-\\

 \hline
\end{tabular}

$^a$ GRBs marked as bold fonts belong to the low-optical-luminosity group, with
separation at $L_{R,1d} \sim 1.4\times 10^{45}$ erg. s$^{-1}$ (see Figure 2).

\end{table}

\begin{figure}
\plotone{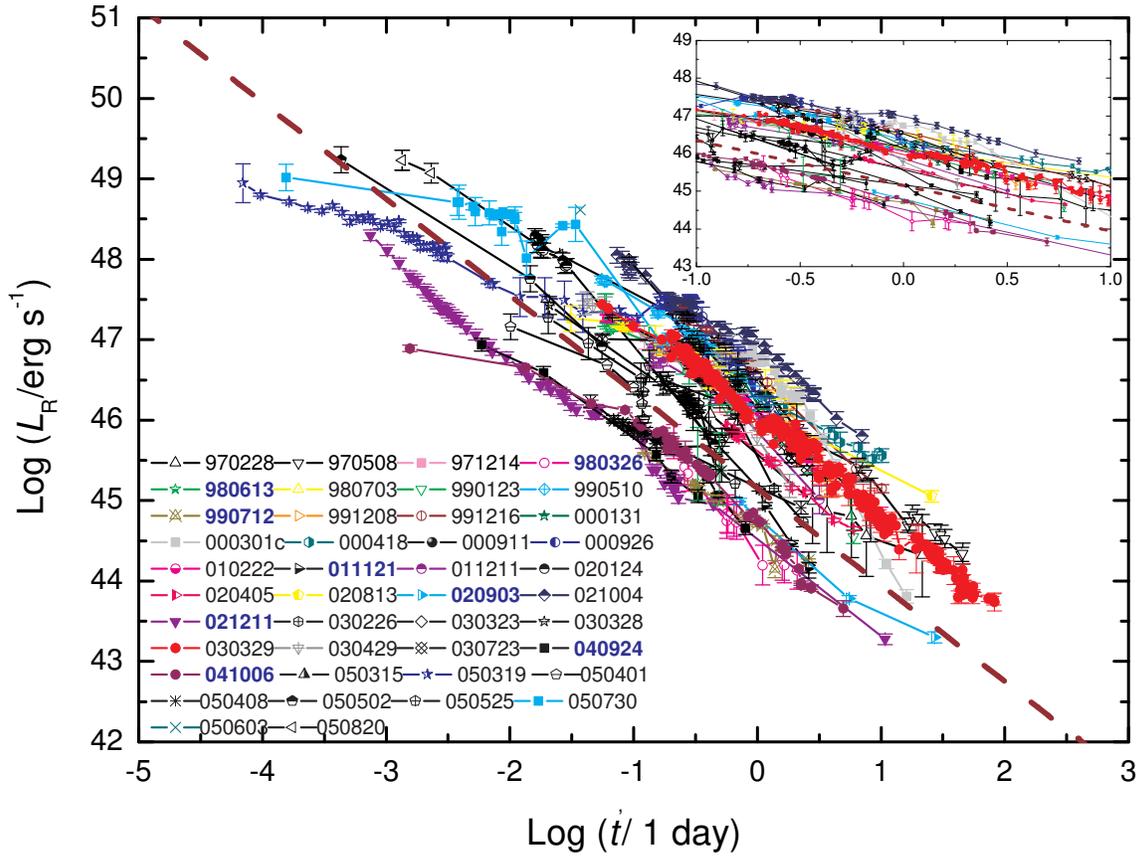} \caption{The R-band light curves ($L_R(t')$ vs. $t'$) in the cosmic
proper rest frame. The dashed line is a division of the two groups of GRBs, $\log
L_R=45.15-1.2\log t^{'}$. The upper inset zooms in the light curves in the time regime
from 0.1 days to 10 days. Those bursts marked with blue color in the figure legend belong
to the dim group. \label{fig1}}
\end{figure}

\begin{figure}
\plotone{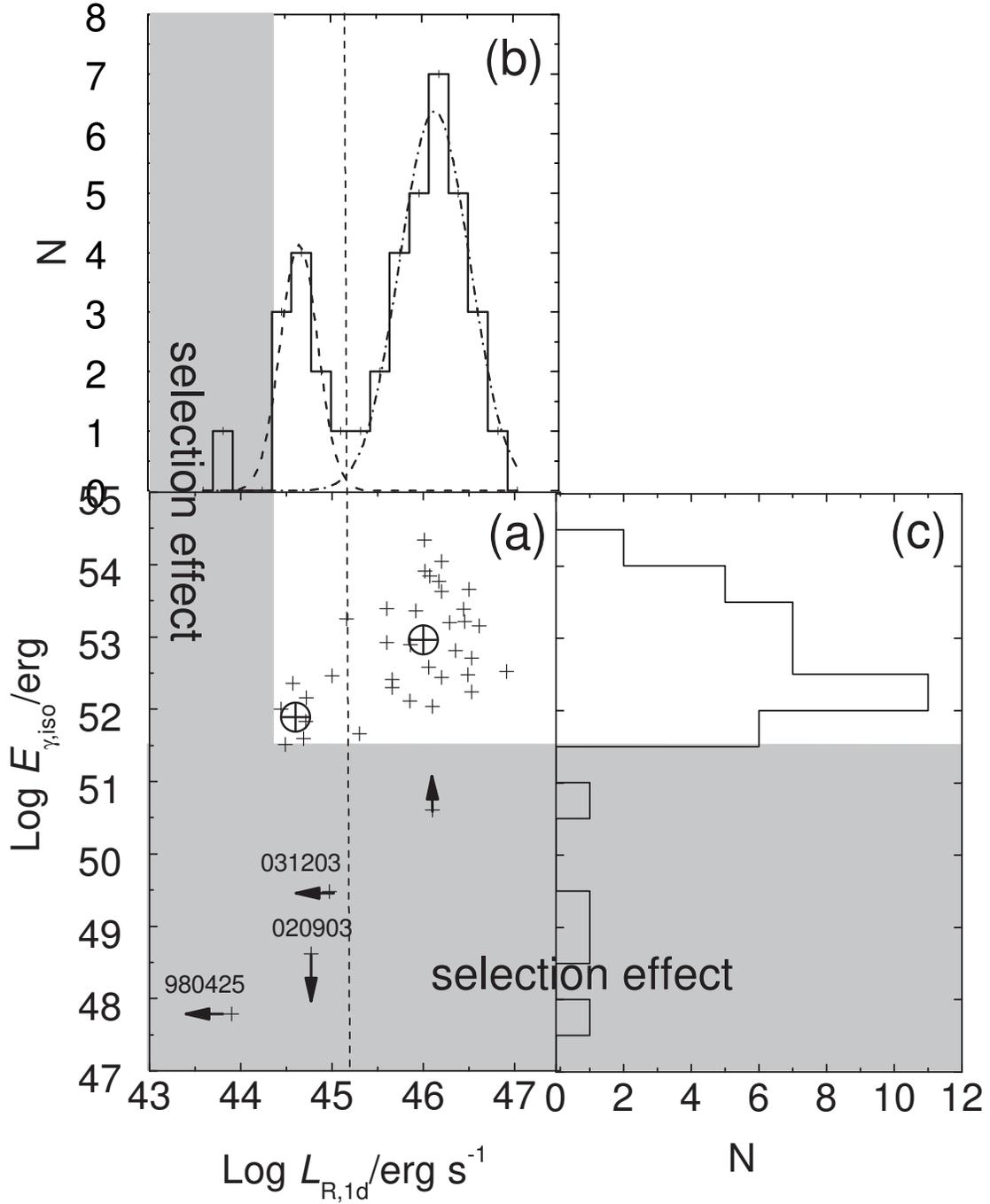} \caption{The 2-dimensional distribution of $L_{R,
1d}$ and $E_{\gamma,\rm iso}$ (panel a), as well as the distributions
of both quantities (panels b and c) for the bursts in our sample. The
significant outliers, GRBs 030226, 970508, and 050408 have been
excluded. The $E_{\gamma,\rm iso}$ has been corrected to the band pass
$20-2000$ keV in the rest frame according to the spectral parameters
of prompt gamma-ray emission. The circled-crosses are the means of the
two quantities for the two groups (excluding those bursts with
limits). The grey area marks the parameter region in which the
flux-threshold selection effect plays a dominant role. The dotted line
in panel (b) is the best fit using a two Gaussian model. The
perpendicular dotted-line is the separation between the dim and the
luminous groups in the two Gaussian model.\label{fig2}}
\end{figure}

\newpage

\begin{table*}[t]
Appended below is the full version of Table 1 with references to the observational data.
It is available in the electronic version in ApJ Letters.

\begin{tabular}{llllllllll}

\hline\hline GRB$^{a}$& $z$ & $\beta(\Delta \beta)$& $A_{\rm V, host}(\Delta A_{\rm V,
host})$ & Ref$^{b}$
\\
\hline

970228&0.695&0.780(0.022)  &0.5 &1;2;2-3\\
970508&0.835 &1.11  &0&4;5;5-6\\
971214&3.42&0.87(0.13) &0.43 (0.08)  &7;8;8-9\\
{\bf 980326}& 1.0 &0.8(0.4) &0&10;10;10-11\\
{\bf 980425}&0.0085 &-&-&12;-;13\\
{\bf 980613}&1.096 &0.60&0.45&14;15;15\\
980703& 0.966&1.013 (0.016)&1.50 (0.11)  &16;17;17-20\\
990123& 1.6004 &0.750 (0.068)&0& 21;22;22-24\\
990510&1.6187 &0.55 &0&25;26;26-28\\
{\bf 990712}&0.434&0.99 (0.02) &0 &25;29;29-30\\
991208& 0.706&0.75 &0 &31;32;32\\
991216& 1.02 &0.60 &0&33;32;32,34\\
000131& 4.5&0.70 &0.18 &35;35;35\\
000301C& 2.03 &0.70 &0.09 &36;37;37\\
000418& 1.118&0.75  &0.96 &38;39;39\\
000911& 1.058&0.724(0.006) &0.39 &40;41;41-42\\
000926& 2.066&1.00(0.18)  &0.18(0.06) &43;44;44\\
010222& 1.477&1.07 (0.09)  &0 &45;46;46\\
{\bf 011121}& 0.36&0.80(0.15)  &0  &47;48;48\\
011211& 2.14&0.56(0.19)&0.08(0.08)  &49;50;51-54 \\
020124& 3.198&0.91 (0.14) &0 &55;55;55-56\\
020405& 0.69&1.43(0.08) &0&57;58;58-59\\
020813& 1.25&0.85(0.07)  &0.14(0.04)&60;61;61-62\\
{\bf 020903}&0.25&-&-&63;-;63\\
021004& 2.335 &0.39 &0.3&64;65;65-66\\
{\bf 021211}& 1.01 &0.69 &0 &67;68;68-70\\
030226& 1.98&0.70(0.03) &0  &71;72;72-73\\
030323& 3.372&0.89(0.04) &$<0.5$  &74;74;74\\
030328& 1.52&-&- &75;-;76-83\\
030329& 0.17&0.5  &0.30(0.03)  &84;85;85-87\\
030429& 2.65&0.75 &0.34 &88;89;89\\
030723&2.10&1.0&0.4&90;90;90\\
{\bf 031203}& 0.105&-&- &91;-;92\\
{\bf 040924}& 0.859&0.70 (0)  &0.16&93;94;94-101\\
{\bf 041006}& 0.716&0.55 &0&102;102;102\\
050315& 1.949&-&-&104;-;105-107\\
050319& 3.24&-&- &108;-;109-115\\
050401& 2.90 &-&-&116;-;117-121\\
050408& 1.24&-&-  &122;-;123-128\\
050502& 3.793& -&-  &129;-;130\\
050525& 0.606&0.97(0.10) &0.25(0.16) &131;132;132-133\\
050730&3.97&-&-&134;-; 134-141\\
050820&2.615&-&-&142;-;143-148\\
 \hline
\end{tabular}

{\bf Notes:}

$^a$ GRBs marked as bold font belong to the low-optical-luminosity group; others belong
to the high-optical-luminosity group.

$^b$ References: three groups of references separated by semicolons are for $z$; $\beta$
and host galaxy extinction; light curve data, respectively. A hyphen is marked when no
reference is available.

\end{table*}
\clearpage

 {\bf References:}

1. Bloom, J. S., Djorgovski, S. G., \& Kulkarni, S. R. 2001, ApJ, 554, 678

2. Galama, T. J., et al. 2000, ApJ, 536, 185

3. Sahu, K. C., et al. 1997, nature, 387, 476

4. Bloom, J. S., et al. 1998, ApJ, 507, L25

5. Galama, T. J., et al. 1998, ApJ, 497, L13

6. Sokolov, V.V., et al. 1998, A\&A, 334, 117

7. Kulkarni, S. R., et al. 1998, nature,  393, 35

8. Wijers, R. A. M. J. \& Galama, T. J. 1999,  ApJ,  523, 177

9. Diercks, A., et al. 1998, ApJ,  503, L105

10. Bloom, J. S., et al. 1999, nature, 401, 453

11. Groot, P. J.,et al.  1998, ApJ, 502, L123

12. Tinney, C., Stathakis, R., Cannon, R., \& Galama, T. J. 1998, IAU 6896, 1

13. Galama, T.J., et al. 1998,nature , 395, 670

14. Djorgovski, S. G., Bloom, J. S., \& Kulkarni, S. R. 2003, ApJ,  591,L13

15. Hjorth, J., et al. 2002, ApJ, 576, 113

16. Djorgovski, S. G., et al. 1998, ApJ, 508, L17

17. Vreeswijk, P. M, et al. 1999,  ApJ, 523, 171

18. Bloom,J. S., et al. 1998, ApJ , 508, L21

19. Castro-Tirado, A. J., et al, 1999, ApJ,  511, L85

20. Frail, D. A., et al. 2003, ApJ,  590,992

21. Kulkarni, S. R., et al. 1999, nature, 398, 389

22. Holland, S., Bj\"{o}rnsson, G.; Hjorth, J., Thomsen, B. 2000, A\&A,  364, 467

23. Fruchter, A. S., et al. 1999, ApJ, 519,L13

24. Castro-Tirado, A. J., et al. 1999, Sci., 283, 2069

25. Vreeswijk, P. M., et al. VLT 2001, ApJ,  546,672

26. Beuermann, K., et al. 1999,  A\&A,  352, L26

27. Harrison, F. A., et al. 1999, ApJ,  523, L121

28. Stanek, K. Z., et al. 1999,  ApJ, 522, L39

29. Sahu, K. C., et, 2000,  ApJ,  540,74

30. Hjorth, J., et al. 2000, ApJ,  534,L147

31. Djorgovski, S. G., et al. 1999, GCN  481

32. Sagar, R., et al. 2000, BASI  28, 15

33. Djorgovski, S. G., et al. 1999, GCN 510

34. Halpern, J. P., et al. 2000, ApJ,  543 697

35. Andersen, M. I., et al. 2000, A\&A,  364, L54

36. Castro, S. M., et al. 2000, GCN,  605

37. Jensen, B. L., et al. 2001, A\&A ,  370, 909

38. Bloom, J. S., et al. 2003, AJ, 125, 999

39. Klose, L., et al. 2000, ApJ ,  545, 271

40. Price, P. A., et al. 2002,  ApJ, 573, 85

41. Masetti, N., et al. 2005,  A\&A, 438, 841

42. Lazzati, D., et al. 2001, A\&A, 378, 996

43. Castro, S. M., et al. 2001, GCN , 851

45. Mirabal, N., et al. 2002, ApJ, 578, 818

46. Stanek, K. Z., et al. 2001, ApJ, 563, 592

47. Garnavich, P. M., et al. 2003,  ApJ, 582, 924

48. Greiner, J., et al. 2003 , ApJ, 599, 1223

49. Gladders, M. et al. GCN,  1209

50. Jakobsson, P. et al. 2003, A\&A, 408,941

51. Holland, S. T., et al. 2002, ApJ, 124, 639

52. Pandey, S. B. et al. 2003,  A\&A, 408, L21

53. Weidong Li 2003, ApJ,   586, L9

54. D. W. Fox et al.  ApJ 586. L5-L8.

55. Hjorth, J., et al. 2003, ApJ, 597, 699

56. Berger, E., et al. 2002, ApJ, 581, 981

57. Price, P. A., et al. 2003 ,ApJ, 589,838

58. Bersier, D., et al. 2003, ApJ, 583,L63

59. Masetti, N., et al. 2003, A\&A, 404, 465

60. Barth, A. J.,et al. 2003,  ApJ, 584, L47

61. Urata, Y., et al.  2003, ApJ, 595, L21

62. Covino, Y. S. et al. 2003, A\&A,  404, L5

63. Soderberg, A. M., et al. 2004, ApJ,  606, 994

64. Giannini, T., et al. 2004, GCN 1678

65. Holland, S. T., et al. 2003, ApJ,  125, 2291

66. Fox, D. W., et al. 2003,  nature, 422, 284

67. Vreeswijk, P.,et al. 2003, GCN , 1785

68. Fox, D. W., et al. 2003,  ApJ,  586, L5

69. Holland, S. T, et al. 2004, ApJ, 128, 1955

70. Li, W. D., et al. 2003, ApJ, 586, L9

71. Greiner, J., et al. 2003, GCN,  1886

72. Klose, S., et al. 2004, ApJ, 128,1942

73. Pandey, S. B., et al. 2004, A\&A,  417, 919

74. Vreeswijk, P. M. et al. 2004, A\&A , 419, 927

75. Martini, P., Garnavich, P., \& Stanek, K. Z. 2003, GCN 1980

76. Gal-Yam, A., et al. 2003, GCN, 1984

77. Fugazza, D., et al. 2003, GCN,  1982

78. Burenin, R., et al. 2003, GCN, 1990

79. Andersen, M. I., et al. 2003,  GCN 1992

80. Martini, P., Garnavich P.  \&  Stanek K.Z. 2003,GCN,1979

81. Bartolini,C. et al. 2003, GCN,  2008

82. Garnavich, P., Martini, P., \& Stanek, K.Z. 2003, GCN, 2036

83. Ibrahimov, M. A.,  et al. 2003,  GCN,   2192

84. Bloom, J. S., Morrell, N., \& Mohanty, S. 2003, GCN,  2212

85. Bloom, J. S., et al .2003, AJ, 127, 252

86. Matheson, T., et al. 2003, A\&A, 599, 394

87. Torii, K., et al. 2003, ApJ, 597,L101

88. Weidinger, M., et al. 2003, GCN,  2215

89. Jak\"{o}bsson P., et el. 2004, A\&A, 427, 785

90. Fynbo, J. P. U. et al. 2004,  ApJ,  609,962

91. Prochaska, J. X., et al. 2003,  GCN, 2482

92. Cobb, B. E., et al. 2004, ApJ, 608, L93

93. Wiersema, K., et al. 2004,  GCN,  2800

94. Soderberg, A. M., et al. 2005, ApJ, 627, 877

95. Fox, D. B., et al. 2004, GCN, 2741

96. Khamitov, I., et al. 2004, GCN, 2740

97. Hu, J. H., et al. 2004,  GCN, 2743

98. Hu, J. H., et al. 2004, GCN, 2744

99. Fynbo, J. P. U., et al. 2004, GCN, 2747

100. Khamitov, I., et al. 2004, GCN,  2749

101. Khamitov, I., et al. 2004, GCN,  2752

102. Price, P. A., et al. 2004, GCN, 2791

103. Stanek, K. Z., et al. 2005, ApJ,  626,  L5

104. Kelson, D. \& Berger, E. 2005, GCN,3101

105. Roming, P. W. A.,et al. 2005, Nature, Submitted

106. Cobb, B. E., et al. 2005, GCN, 3104

107. Cobb, B. E., et al. 2005, GCN, 3110

108. Fynbo, J. P. U., et al. 2005, GCN, 3136

109. Yoshioka, T.,  et al. 2005 , GCN, 3120

110. Torii, K., et al. 2005, GCN, 3121

111. Sharapov, D., et al. 2005, GCN, 3124

112. Misra, K, et al. 2005, GCN, 3130

113. Kiziloglu, U., et al. 2005, GCN, 3139

114. Sharapov, D., et al. 2005, GCN, 3140

115. Greco, G., et al.  2005, GCN, 3142

116. Fynbo,J. P. U., et al. 2005, GCN, 3176

117. McNaught, R., et al. 2005, GCN, 3163

118. D'Avanzo, P., et al. 2005, GCN, 3171

119. Kahharov, B., et al. 2005, GCN, 3174

120. Misra, K.,  et al. 2005, GCN, 3175

121. Greco, B., et al. 2005, GCN, 3319

122. Berger, E., Gladders, M., \& Oemler, G. 2005, GCN,  3201

123. Wiersema, K., et al. 2005, GCN, 3200

124. de Ugarte A., et al. 2005, GCN, 3199

125. Milne, P. A., et al. 2005, GCN, 3258

126. Curran, P., et al. 2005, GCN, 3211

127. Aslan, Z., et al. 2005, GCN, 3198

128. Nysewander, M., et al. 2005, GCN, 3213

129. Prochaska,J. X., et al. 2005, GCN, 3332

130. MirabalN., et al. 2005, GCN 3363

131.Foley,R. J., et al. 2005, GCN,  3483

132. Blustin, A. J., et al. 2005, ApJ, in press (astro-ph/0507515)

133. Torii,K. \& BenDaniel, M. 2005,GCN, 3470

134. Holman, M., Garnavich, P. \& Stanek,  K. Z. 2005, GCN 3716

135. Sota, A. et al. 2005, GCN, 3705

136. Burenin, R. et al. 2005, GCN, 3718

137. Klotz, A., Boer, M., \& Atteia, J. L. 2005,GCN, 3720

138. Damerdji, Y., et al. 2005, GCN, 3741

139. D'Elia, V.  2005, GCN, 3746

140. Bhatt, B. C. et al. 2005, GCN, 3775

141. Kannappan, S. et al. 2005, GCN, 3778

142. Ledoux, C. et al. 2005, GCN, 3860

143. Fox, D, B. et al. 2005, GCN, 3829

144. Cenko, S. B.,et al. 2005, GCN,3834

145. Bikmaev, I.  et al. 2005, GCN, 3853

146. MacLeod, C. et al. 2005, GCN, 3863

147. Khamitov I. et al. 2005, GCN, 3864

148. Aslan, Z. et al. 2005, GCN, 3896


\begin{thebibliography}{}
\bibitem [Ashman et al. 1994]{KMM}Ashman, K. M., Bird, C. M., \& Zepf,
S. E. 1994, AJ, 108, 2348
\bibitem [Barthelmy et al. 2005]{Barthelmy05}Barthelmy, S. D., et
al. 2005, Nature, 438, 994
\bibitem [Berger et al. 2002]{Berger02}Berger, E., et al. 2002, ApJ, 581, 981
\bibitem [Berger, Kulkarni, \& Frail 2003]{Berger03}Berger, E.,
Kulkarni, S. R., \& Frail, D. A. 2003, ApJ, 590, 379
\bibitem [Berger et al. 2005]{Berger05}Berger, E., et al. 2005, Nature, 438, 988
\bibitem [Bj\"{o}rnsson et al. 2004]{Bjornsson04}Bj\"{o}rnsson,
G., Gudmundsson, E. H., \& J\'{o}hannesson, G. 2004, ApJ, 615, L77
\bibitem [Bloom et al. 1999]{Bloom99}Bloom, J. S., et al. 1999, Nature, 401, 453
\bibitem [Bo\"{e}r \&  Gendre 2000]{Boer00}Bo\"{e}r, M. \&  Gendre,
B. 2000, A\&A, 361, L21
\bibitem [Chincarini et al. 2005]{Chincarini05}Chincarini, G., et al.
2005, ApJ, submitted
\bibitem [Costa et al. 1997]{Costa97}Costa, E., et al. 1997, Nature, 387, 783
\bibitem [Dai \& Lu 2001]{Dai01}Dai, Z. G. \& Lu, T. 2001, A\&A,367, 501
\bibitem [Fox et al. 2003]{Fox03}Fox, D. W., et al. 2003, Nature, 422, 284
\bibitem [Fox et al. 2005]{Fox05}Fox, D. B., et al. 2005, Nature, 437, 845
\bibitem [Frail 1997]{Frail97}Frail, D. A., et al. 1997, Nature, 389, 261
\bibitem [Frail et al. 2001]{Frail01}Frail, D. A., et al. 2001, ApJ, 562, L55
\bibitem [Fruchter et al. 2001]{Fruchter01}Fruchter, A., Krolik,
J. H., \& Rhoads, J. E.,  2001, ApJ, 563, 597
\bibitem [Fukugita et al. 1995] {Fukugita95}Fukugita, M., Shimasaku,
K., \& Ichikawa, T. 1995, PASP, 107, 945
\bibitem [Fynbo et al. 2001]{Fynbo01}Fynbo, J. U., et al. 2001, A\&A,
369, 373
\bibitem [Galama \&  Wijers 2001]{Galama01}Galama, T. J., \&  Wijers,
R. A. M. J. 2001, ApJ, 549, L209
\bibitem [Galama et al. 1998]{Galama98}Galama, T. J., et al. 1998,
Nature, 395, 670
\bibitem [Gehrels et al. 2005]{Gehrels05}Gehrels, N., et al. 2005,
Nature, 437, 851
\bibitem [Gendre \& Bo\"{e}r 2005]{Gendre05}Gendre, B. \& Bo\"{e}r,
M. 2005, A\&A, 430, 465
\bibitem [Groot et al. 1998]{Groot98} Groot, P. J., et al. 1998, ApJ,
493, L27
\bibitem [Guetta et al. 2004]{Guetta04}Guetta, D., et al. 2004, ApJ,
615, L73
\bibitem [Hjorth et al. 2003]{Hjorth03}Hjorth, J., et al. 2003, Nature, 423, 847
\bibitem [Hjorth et al. 2005a]{Hjorth05a}Hjorth, J., et al. 2005a,
ApJ, 630, L117
\bibitem [Hjorth et al. 2005b]{Hjorth05b}Hjorth, J., et al. 2005b,
Nature, 437, 859
\bibitem [Holland et al. 2004]{Holland04} Holland, S. T, et al., 2004,
ApJ, 128, 1955
\bibitem [Jakobsson et al. 2004]{Jakobsson04} Jakobsson, P., et al.
2004, ApJ, 617, L21
\bibitem [Kouveliotou et al. 1993]{Kouveliotou93}Kouveliotou, C., et
al. 1993, ApJ, 413, L101
\bibitem [Lloyd-Ronning \& Zhang 2004]{Lloyd04}Lloyd-Ronning,
N. M. \& Zhang, B. 2004, ApJ, 613, 477
\bibitem [MacFadyen \& Woosley 1999]{MacFadyen99}MacFadyen, A. I. \&
Woosley, S. E. 1999, ApJ, 524, 262
\bibitem [Malesani et al. 2004]{Malesani04}Malesani, D., et al. 2004, ApJ, 609, L5
\bibitem [M\'{e}sz\'{a}ros \& Rees 1997]{Meszaros97}M\'{e}sz\'{a}ros,
P. \& Rees, M. J. 1997, ApJ, 476, 232
\bibitem [M\'{e}sz\'{a}ros 2002]{Meszaros02}M\'{e}sz\'{a}ros, P. 2002, ARA\&A, 40, 137
\bibitem [Metzger et al. 1997]{Metzger97}Metzger, M. R., et al. 1997, Nature, 387, 879
\bibitem [Nardini et al. 2005]{Nardini05} Nardini, N., et al. 2005, A\&A, submitted (astro-ph/0508447)
\bibitem [Norris 2002]{Norris02}Norris, J. P. 2002, ApJ, 579, 386
\bibitem []{}Pedersen, H., et al. 1998, ApJ, 496, 311
\bibitem [Pei 1992]{Pei92}Pei, Y. C. 1992, ApJ, 395, 130
\bibitem [Piran 2005]{Piran05}Piran, T. 2005, Rev. Mod. Phys., 76, 1143
\bibitem [Rhoads 1999]{Rhoads99}Rhoads, J. E., 1999, ApJ, 525, 737
\bibitem [Rol et al. 2005]{Rol05}Rol, E., et al. 2005, ApJ, 624, 868
\bibitem [Roming et al. 2005]{Roming05} Roming, P. W. A., et al.
2005, ApJ, submitted (astro-ph/0509273)
\bibitem [Rossi, Lazzati, \& Rees 2002]{Rossi02}Rossi, E., Lazzati,
D., \& Rees, M. J. 2002, MNRAS, 332, 945
\bibitem [Sari et al. 1998]{Sari98}Sari, R., Piran, T., \& Narayan,
R. 1998, ApJ, 497, L17
\bibitem [Schlegel et al.  1998] {Schlegel98}Schlegel, D. J.,
Finkbeiner, D. P, \& Davis, M. 1998, ApJ, 500, 525
\bibitem [Schmidt 2001]{Schmidt01}Schmidt, M. 2001, ApJ, 552, 36
\bibitem [Soderberg et al. 2004]{Soderberg04}Soderberg, A. M., et al.
2004, Nature, 430, 648
\bibitem [Stanek et al. 2003]{Stanek03}Stanek, K. Z., et al.  2003,
ApJ, 591, L17
\bibitem [Thomsen, et al. 2004]{Thomsen04}Thomsen, B., et al.  2004,
A\&A 419, L21
\bibitem [van Paradijs et al. 1997]{Van97} van Paradijs, et al. 1997,
Nature, 386, 686
\bibitem [Villasenor et al. 2005]{Villasenor05}Villasenor, J. S, et
al. 2005, Nature, 437, 855
\bibitem [Waxman \& Draine 2000]{Waxman00}Waxman, E. \& Draine,
B. T. 2000, ApJ, 537, 796
\bibitem [Zhang et al. 2005]{Zhang05}Zhang, B., et al. 2006, ApJ, in press (astro-ph/0508321)
\bibitem [Zhang \& M\'{e}sz\'{a}ros 2002]{Zhang02}Zhang, B. \&
M\'{e}sz\'{a}ros, P. 2002, ApJ, 571 876

\bibitem [Zhang \& M\'{e}sz\'{a}ros 2004]{Zhang04}Zhang, B. \&
M\'{e}sz\'{a}ros, P. 2004, Int. J. Mod. Phys. A, 19, 2385
\end{thebibliography}
\end{document}